\documentstyle[epsf]{elsart}

\begin{document}

\begin{frontmatter}

\title{Phase Coherence in Chaotic Oscillatory Media}
\author[ufrgs]{Leonardo G. Brunnet} and
\author[spec]{Hugues Chat\'e}

\address[ufrgs]{Instituto de F\'{\i}sica,
Universidade Federal do Rio Grande do Sul,
Caixa Postal 15051, 91501-970 Porto Alegre (RS), Brazil}

\address[spec]{CEA --- Service de Physique de l'Etat Condens\'e,
Centre d'Etudes de Saclay, 91191 Gif-sur-Yvette, France}

\begin{abstract}
Collective oscillations of lattices of locally-coupled chaotic R\"ossler
oscillators are studied with regard to the dynamical scaling 
of their phase interfaces. Using analogies with the
complex Ginzburg-Landau  and the Kardar-Parisi-Zhang equations, 
we argue that phase coherence should be lost in the infinite-size 
limit. Our numerical results, however, indicate possible discrepancies 
with a Langevin-like description using an effective white-noise term.
\end{abstract}

\end{frontmatter}

Spatially-extended, extensively-chaotic dynamical systems with local 
interactions generically exhibit some collective coherence emerging out
of strong local chaos which seems to persist even in the infinite-size
limit.  This long-range order in far-from-equilibrium, 
deterministic systems often takes the form of a simple, effectively
low-dimensional, temporal evolution of spatially-averaged quantities. 
Usually referred to as non-trivial collective behavior, this
phenomenon has been studeid mostly in discrete-time lattice systems
such as coupled map lattices and cellular automata \cite{NTCB}.
In a recent paper, though, a continuous-time model was investigated in this 
context \cite{ROSS}. 
It concluded from numerical experiments that two-dimensional
lattices of diffusively-coupled chaotic R\"ossler systems may 
show collective 
oscillations, or long-range rotating order (LRRO). In this Paper,
we look at this result in a new light, by 
making use of the properties of the Kardar-Parisi-Zhang equation (KPZE)
\cite{KPZ}, a universal model for fluctuating interfaces.

\section{Motivation}

A lattice of R\"ossler systems coupled to their nearest neighbors
by diffusion can be schematically written:
\begin{equation}
\label{lattice}
C^i_{t+\tau} = (1-\varepsilon) {\cal F}_{\tau}(C^i_t) +
 \frac{\varepsilon}{\cal N}
\sum_{j\in {\cal V}_i} {\cal F}_{\tau}(C^j_t) \;,
\end{equation}
where $C^i$ is the three-component vector sitting at site $i$, $\varepsilon$
is the coupling strength, $\tau$ is the interval between coupling times,
${\cal N}$ is the number of nearest-neighbors, ${\cal V}_i$ is the 
neighborhood of site $i$, and ${\cal F}_{\tau}(C^i_t)$
represents the state of $C_i^t$ after evolution under the R\"ossler flow 
${\cal R}$ during a time $\tau$. In other words, dropping the subscript $i$:
\begin{equation}
{\cal F}^{\tau}(C^t) = C^t + \int_{t}^{t+\tau} \! dt' \: \dot{C} \;.
\end{equation}

\begin{figure}
\vspace{-3.5cm}
\centerline{
\epsfxsize 12cm
\epsffile{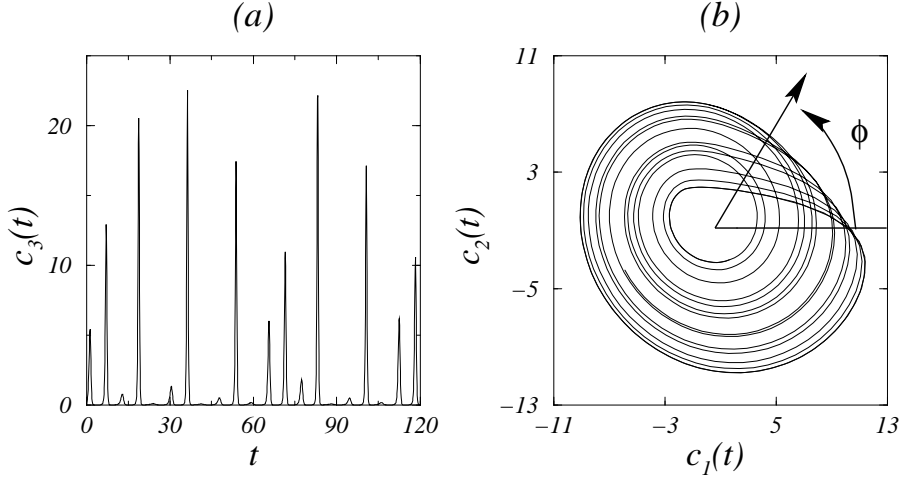}
}
\vspace{-0.3cm}
\caption{Chaotic dynamics of the R\"ossler system (\protect\ref{rossler})
with $a=b=0.2$ and $c=5.7$.
(a): time series of $c_3$.
(b): attractor in the $(c_1,c_2)$ plane and definition of phase.}
\label{f1}
\end{figure}

The R\"ossler model possesses remarkable properties. It is usually written 
as a three-variable, first-order, ordinary differential system:
\begin{equation}
\label{rossler}
\dot{C}=\pmatrix{\dot{c_1} \cr \dot{c_2} \cr \dot{c_3} \cr} = 
\pmatrix{-c_2-c_3 \cr c_1+ac_1 \cr b+c_1c_3-cc_3 \cr } \equiv {\cal R}(C)
\end{equation}
where $a$, $b$, and $c$ are real parameters. Increasing $c$
while keeping $a$ and $b$ fixed (for example $a=b=0.2$ as in \cite{ROSS}),
system (\ref{rossler}) undergoes a Hopf bifurcation followed by a 
cascade of subharmonic bifurcations eventually leading to chaos. 
For these parameter values, the chaotic attractor is 
characterized by  $c_3$  peaks of irregular amplitude but almost
perfectly-defined frequency  (Fig.~\ref{f1}a). 
Similarly, trajectories in the $(c_1,c_2)$
plane are cycles of irregular amplitude but with a well-defined period
 (Fig.~\ref{f1}b).
This allows the definition of ``phase'' and ``amplitude'' variables, either
simply by using the $(c_1,c_2)$ plane with an origin set in the middle of
the attractor (Fig.~\ref{f1}b), or by more sophisticated methods.
One can thus speak of a ``chaotic oscillator''  \cite{PICK}.
Picturing the R\"ossler system as an oscillator translates the
problem of the LRRO observed in \cite{ROSS} into a phase coherence,
or synchronization, problem for a chaotic oscillatory medium\footnote{Note 
that the remarkable phase coherence
of the R\"ossler model was recently studied and quantified in \cite{PICK}, 
where the possibility of (exact) synchronization of these systems
was evidenced.}.

In \cite{ROSS}, the chaotic oscillatory behavior of lattices
of R\"ossler systems of the type (\ref{lattice}-\ref{rossler}) 
was also used to draw an analogy with the 
complex Ginzburg-Landau equation (CGLE), the generic nonlinear 
partial differential equation 
describing an oscillatory medium near a Hopf bifurcation \cite{CGL}. 
The CGLE reads:
\begin{equation}
\partial_t A = A +(1+i\alpha) \nabla^2 A - (1+i\beta) |A|^2 A
\end{equation}
where $A$ is a complex field, $\alpha$ and $\beta$ are real parameters.
It was argued that the LRRO observed can be well accounted for by
a CGLE with parameters corresponding to a regime where homogeneous 
oscillations are (linearly) stable, with some residual ``effective'' noise.
In that analogy, the complex plane roughly corresponds to the $(c_1,c_2)$
plane of the R\"ossler variables.
On general grounds, one expects the soft phase modes of the noisy stable
CGLE
to be described at large scales by the Kardar-Parisi-Zhang equation 
(KPZE)\footnote{At least when no zeroes of the complex field $A$ 
are present and when the above-mentioned effective noise is 
delta-correlated.},
a stochastic model for the kinetic roughening of 
fluctuating interfaces \cite{KPZ} which reads:
\begin{equation}
\label{kpz}
\partial_t h = \nu \nabla^2 h + \frac{\lambda}{2}
(\nabla h)^2 +\eta ({\bf x},t)
\end{equation}
where $h$ is a real field, $\nu$ and $\lambda$ are real parameters,
and $\eta({\bf x},t)$ is an uncorrelated white noise with zero mean 
and correlators:
\begin{equation} 
\langle\eta({\bf x}, t) \eta({\bf x}', t^\prime)\rangle
=2 D\delta(t-t^\prime) \delta({\bf x}-{\bf x}^\prime).
\label{noise}
\end{equation}
In (\ref{kpz}), $h$ is the height of an interface and 
takes values on the entire real axis. In the present context,
$h$ represents the angular (or phase) argument of
the complex field $A$ (or the phase $\phi$ of the R\"ossler oscillators)
followed by continuity in space and time
from some arbitrary initial value. In this representation,
the phase coherence problem described above
 can be restated in terms of the roughness of the phase interface:
if the interface is rough (its mean square width diverges in the
infinite-size infinite-time limit), then no phase coherence exists.

The KPZE has gained considerable importance because many ``microscopic''
models share its non-trivial scaling properties, and also because
some analytical results have been obtained \cite{KPZ}. 
Among those, one is crucial
here: interfaces governed by the KPZE are always rough for space dimensions
$d\le 2$. Assuming the validity of the KPZE to describe the large-scale
properties of lattices of coupled R\"ossler oscillators, this result
is at odds with the conclusion reached in \cite{ROSS}. Thus, either
the numerical results of \cite{ROSS} were too limited to reveal the loss
of phase coherence in the infinite-size, infinite-time limit, or 
the KPZE is not the correct stochastic equation.
In the latter case,
the discrepancy probably lies in the properties of the 
``effective noise''\footnote{The presence of the diffusive term 
is guaranteed by the coupling function
in (\ref{lattice}), and that of the quadratic nonlinear term is insured
by the relevance of the CGLE demonstrated in \cite{ROSS}. 
Indeed, for the CGLE,
the $(\nabla h)^2$ term represents the quadratic dependence of frequency on
wavenumber implied by the nonlinear term  $|A|^2 A$.}.
 
As a matter of fact, there is no {\it a priori} reason for
the KPZE to be the relevant large-scale description of the phase dynamics
of coupled R\"ossler oscillator lattices. But the ``universality class''
of the KPZE has been shown to be remarkably large. In particular, recent
findings show it to include the phase interface dynamics generated
by the CGLE in its so-called ``phase turbulence'' regime  \cite{TURBPHA}. 
In this spatiotemporally chaotic regime, 
there are no zeroes of the complex field $A$,
and a fluctuating continuous phase interface can always be 
defined\footnote{In fact, it was also concluded in \cite{TURBPHA} that phase 
turbulence always eventually breaks down, leading to the occurrence of 
zeroes of $A$. But there exist large portions of the parameter plane 
($\alpha, \beta$) in which statistically-steady phase turbulence can be 
observed on very large scales, so that, in practice, breakdown is never 
observed.}.
To some extent, this case might appear very similar to the chaotic
oscillatory media formed by lattices of R\"ossler systems; thus,
one would expect the KPZE to be relevant. On the other hand, as recalled 
above, chaos in the R\"ossler system possesses some very specific features
that might possibly produce an ``effective noise'' with peculiar 
properties.

Given the space-time evolution of some interface, there exists a rather
well-established procedure to determine whether the KPZE is a relevant
large-scale description \cite{KRUG}. 
In the following, we briefly recall this procedure and follow it to
investigate the coherent oscillations in lattices of coupled R\"ossler
systems from the angle of the phase interface dynamics they produce.

\section{Experimental procedure}

We have performed numerical experiments of system 
(\ref{lattice}-\ref{rossler}) and studied the scale-invariance properties 
of the interface constructed from the ``phase'' of each of the R\"ossler
oscillators. The calculations were done in one space dimension,
mostly for numerical convenience --- if phase coherence 
is to be broken, it should be easier to observe for $d=1$ --- 
but also because many exact results are known in this case for the KPZE.
Scale-invariance relies on the scaling assumption
$h(\ell {\bf x}, {\ell}^z t) = \ell^\zeta h({\bf x},t)$,
where $\ell$ is a similarity factor, $\zeta$ and $z$ are
(respectively) the roughness and dynamical exponents. For interfaces 
governed by the KPZE, exponents are exactly known for $d=1$: $z=\frac{3}{2}$
and $\zeta=\frac{1}{2}$. All scaling laws given in the following are 
for these values.

In practice, one usually considers global quantities such as the mean square
width of the interface:
\begin{equation}
w^2(t) = \left\langle \left(h({\bf x},t)-\langle h\rangle_{\bf x} \right)^2 
\right\rangle_{\bf x} \; ,
\end{equation}
where $\langle\dots\rangle_{\bf x}$ denotes space average. 
For a system of finite-size $L$, the width of an initially flat interface 
grows and saturates to a size-dependent mean value:
\begin{equation}
\label{saturated}
w^2(L,t\rightarrow\infty) = \frac{D}{24\nu}\, L \; .
\end{equation}
For an infinitely large system, $w^2$ grows indefinitely:
\begin{equation}
\label{nlgrowth}
w^2(L\rightarrow\infty, t) \simeq 0.4\times \left(\frac{D^2}{4\nu^2}
\lambda\, t\right)^{2/3} \;. 
\end{equation}
For the KPZE, this growth phase actually takes place only beyond 
certain crossover scales \cite{LHT}: 
\begin{equation}
\label{crossover}
L_c\simeq \frac{152}{g} \;\;\;{\rm and}\;\;\; t_c\simeq \frac{252}{\nu g^2} 
\;\;\; {\rm with}
\;\;\; g=\frac{\lambda^2 D}{\nu^3}\; ,
\end{equation}
before which another scaling 
is observed, because the  nonlinearities are not yet effective. 
The ``linear'' growth phase is characterized by a growth exponent 
$\beta=(2-d)/4=1/4$ for $d=1$. One expects:
\begin{equation}
\label{lingrowth}
w^2(L\rightarrow\infty, t) = \frac{D}{\sqrt{2\pi\nu}} \, t^{1/2}  \;.
\end{equation}

Relations (\ref{saturated}), (\ref{nlgrowth}), and (\ref{lingrowth})
allow one to check dynamical scaling  and to estimate whether
the measured exponents are consistent with those of the KPZE. In addition,
the measure of the numerical prefactors of the scaling laws can lead
to a determination of the effective 
parameters $\nu$, $D$, and $\lambda$ of the corresponding KPZE and of the 
crossover scales $L_c$ and $t_c$, provided that $\lambda$ is determined
independently. This is usually achieved by measuring the changes in the 
velocity of the interface 
$v={\rm d}\langle \phi\rangle_{\rm x}/{\rm d}t$ 
when it is submitted to a tilt $q=2\pi n/L$, where $n$ is an integer ``winding
number'', using the relation \cite{KRUG}:
\begin{equation}
\lambda =\left. \frac{{\rm d}^2 v}{{\rm d}\,q^2}\right|_{q=0} \;.
\label{tilt}
\end{equation}

\begin{figure}
\vspace{0cm}
\centerline{\hspace{1cm}
\epsfysize 8cm
\epsffile{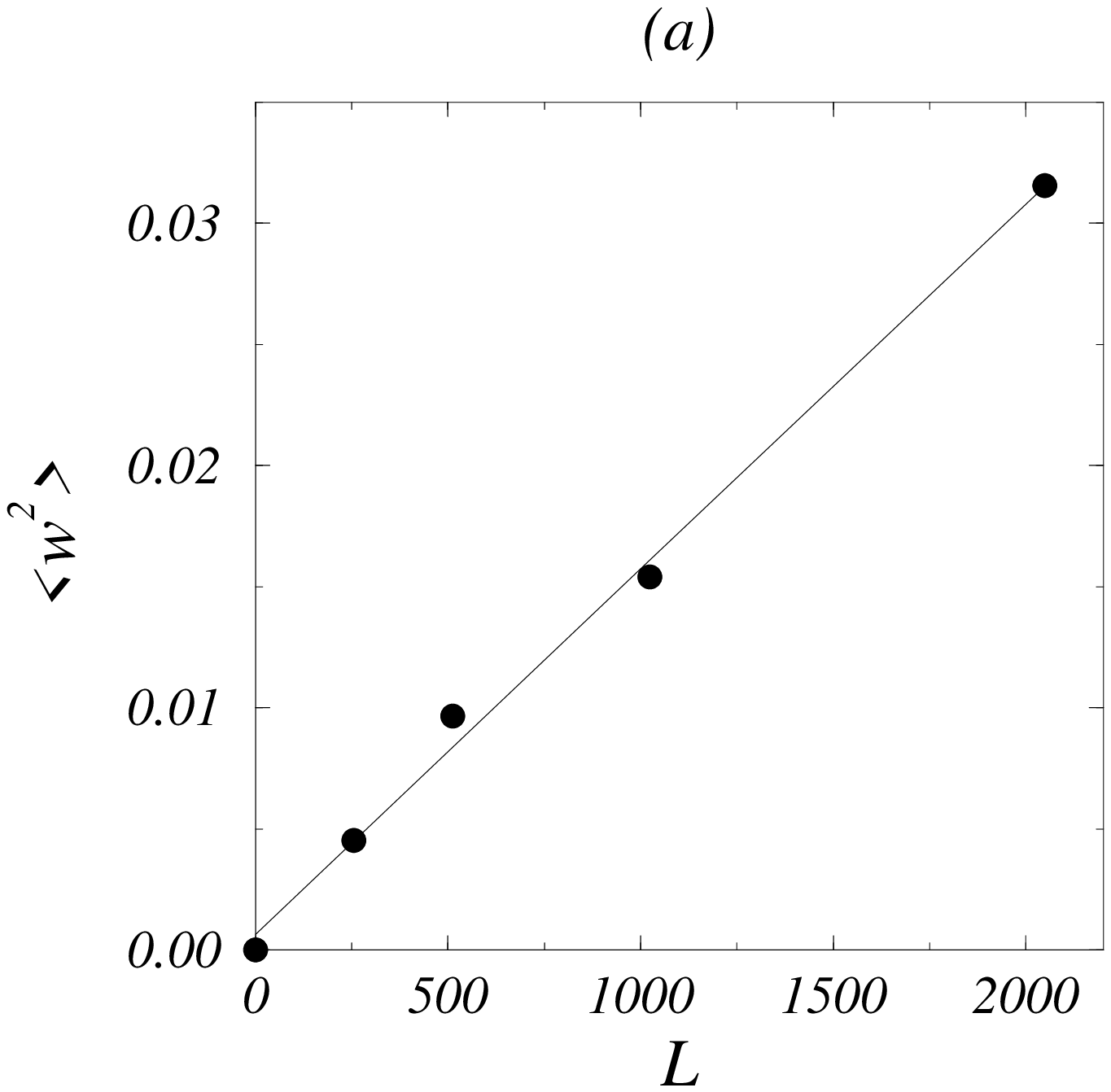}
\hspace{-3cm}
\epsfysize 8cm
\epsffile{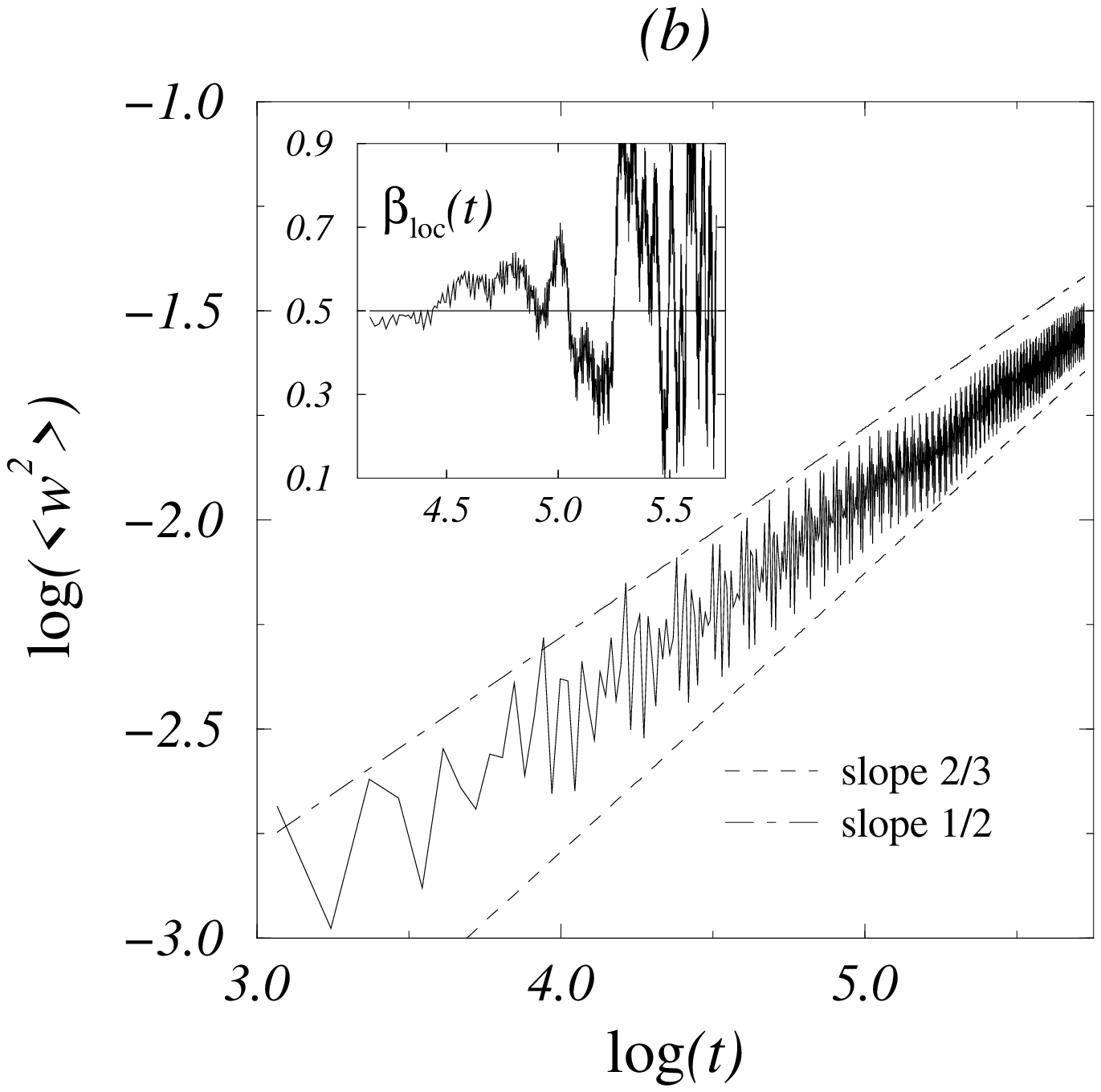}
}
\vspace{-1cm}

\caption{Chain of diffusively coupled R\"ossler systems. (a): the 
ensemble-averaged, saturated,  square mean width scales with system size $L$,
with a slope $D/24\nu \simeq 1.5\times 10^{-5}$. 
(b): in the growth regime, a system of $2^{18}$ oscillators seems to remain 
in the linear regime (single run). 
Lines of slope $1/2$ (linear regime) and $2/3$
(KPZE nonlinear regime) are shown. 
The insert shows the time evolution of the ``local growth exponent'' 
$\beta_{\rm loc}(t)$ calculated over a time-window $\Delta t = 15000$. 
We cannot rule out the beginning of a crossover from the value $1/2$ to some
larger value, but the data is too noisy to conclude.}
\label{f2}
\end{figure}

\section{Chain of R\"ossler systems with purely diffusive coupling}

We first consider a chain of R\"ossler systems with periodic boundary
conditions, as defined by (\ref{lattice}-\ref{rossler}). 
The coupling strength and the coupling interval are set
to $\varepsilon=1/6$ and $\tau=1.8$, values  
which insure the quasi-continuity of the medium and thus of the phase
interface. The parameters of the 
R\"ossler systems themselves have the same values as in \cite{ROSS}:
$a=b=0.2$ and $c=5.7$.
Starting from random initial conditions far from the center of the 
R\"ossler attractor,
all oscillators quickly reach nearby values,
yielding an initially quasi-flat interface.

For system sizes $L\le 2048$, saturation of the width of the interface could
be observed, as well as the linear scaling with system size
(Fig.~\ref{f2}a). Regarding the growth of $w$ in a large system,
even for the largest size ($L=2^{18}$) 
and the longest times ($t\sim 5\times 10^5$) considered, only the linear
regime ($w^2\sim t^{1/2}$) seems to be observed (Fig.~\ref{f2}b).
From these results, we also estimate $D/24\nu \simeq 1.5\times 10^{-5}$
and $D/\sqrt{2\pi\nu} \simeq 4\times 10^{-5}$, and thus 
$\nu\simeq 8\times 10^{-2}$, $D\simeq 3\times 10^{-5}$.

To estimate $\lambda$,
we performed tilt experiments, preparing initial conditions with a prescribed
winding number $n$, and measuring the velocity of the phase interface.
Following Eq.~(\ref{tilt}),
only the small $q$ behavior is of interest. However, for too small tilts,
the variations of $v$ are not numerically measurable. Consequently,
a reliable measure of $\lambda$ is very difficult. Our results,
obtained for moderate tilts, give $\lambda\simeq -2.7$. This is 
in rough agreement with  \cite{ROSS}, since, 
in the stable CGLE context, $\lambda=2(\beta-\alpha)$ with the parameters
estimated at $\alpha\simeq 0.66$ and $\beta\simeq -1.06$

Gathering these results together, we obtain the following estimates:
$g\sim 0.4$, $L_c\sim 350$, and $t_c\sim 2\times 10^{-4}$.
Clearly, there is a contradiction
between these estimates and the recorded behavior,
which remained in the linear regime well beyond these scales.
We will come back to this point in the discussion.

\begin{figure}
\vspace{0cm}
\centerline{\hspace{1cm}
\epsfysize 8cm
\epsffile{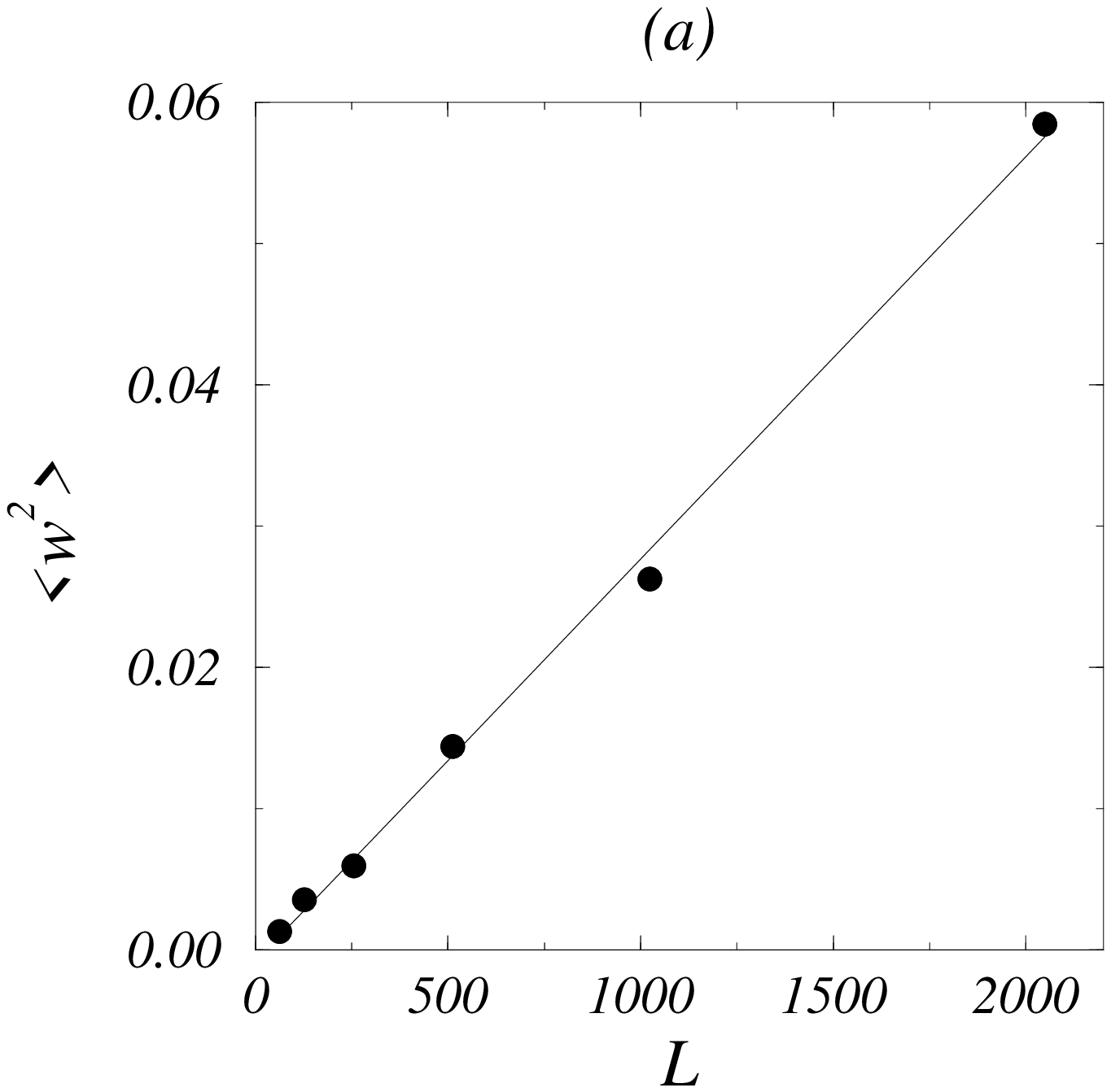}
\hspace{-3cm}
\epsfysize 8cm
\epsffile{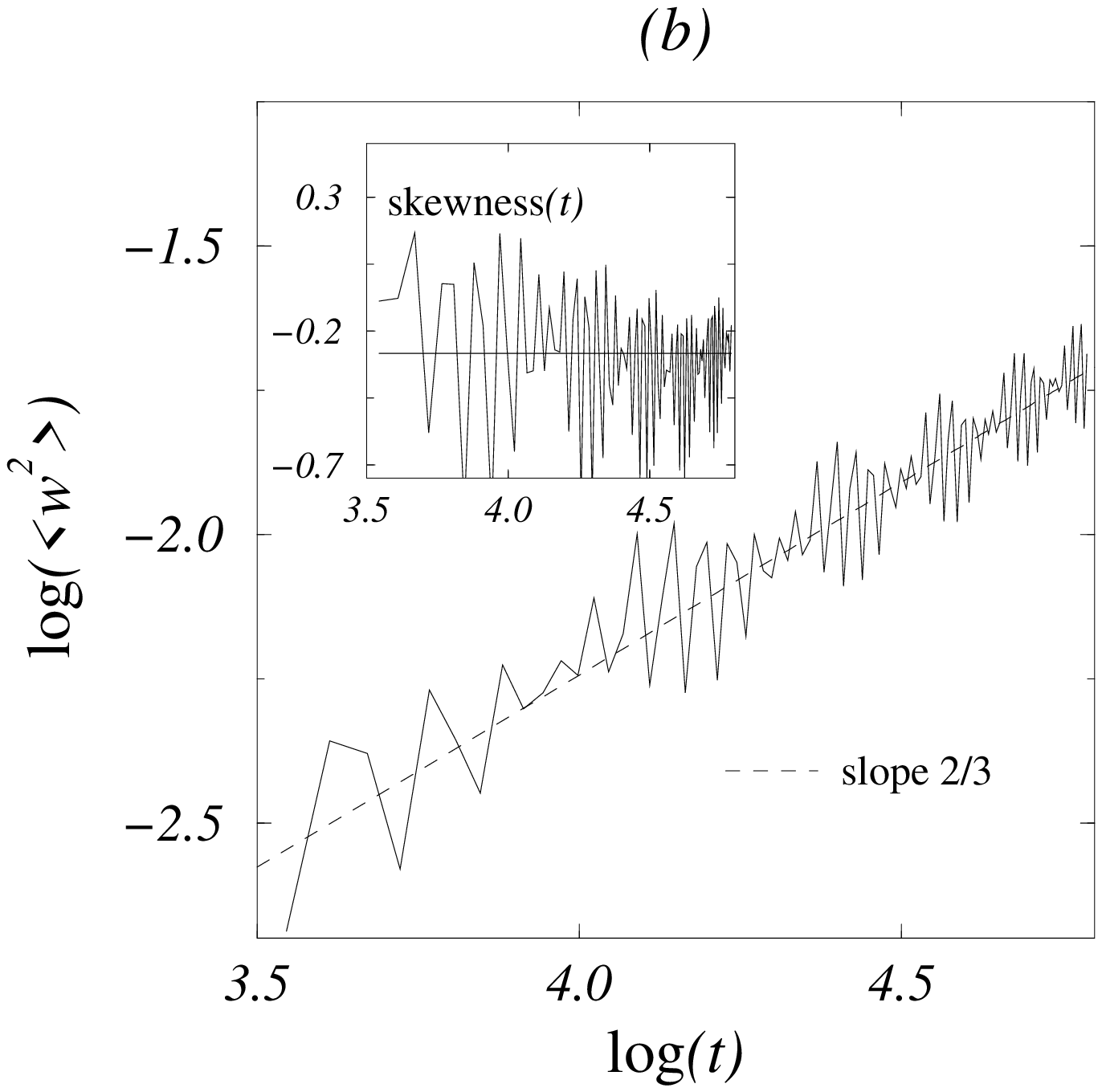}
}
\vspace{-1cm}

\caption{Chain of R\"ossler systems with diffusive-dispersive coupling. 
(a): the ensemble-averaged, saturated,  square mean width scales 
with system size $L$, with a slope $D/24\nu \simeq 2.85 \times  10^{-5}$. 
(b): in the growth regime, a system of $2^{18}$ oscillators quickly reaches the
KPZE nonlinear behavior, characterized by a growth exponent $\beta=2/3$
and a skewness of mean value $\simeq -0.285$ (insert, solid line).}
\label{f3}
\end{figure}

There is one remarkable
fact in the above results: the largest widths reached during our calculations
are always very small (at most of the order of $2\pi$). ``Roughening'' is thus
extremely weak in this system, even though the natural extrapolation of 
our numerical results is that, indeed, phase coherence should be lost in the
infinite-size, infinite-time limit. In the next section, we consider
the same system but with a modified coupling designed to increase the
roughening of the phase interface.

\section{Chain of R\"ossler systems with diffusive-dispersive coupling}
\label{s4}

For the two-dimensional lattice of \cite{ROSS},
the effective CGLE was found to be in a regime where
the spatially-homogeneous solution $A=\exp (-i\beta t)$ is linearly stable.
One way of bringing the effective CGLE into an intrinsically chaotic
regime ---and thus, hopefully, to strengthen roughening--- 
is to increase $\alpha$, so as to be in a phase turbulence regime 
(which is reached when $1+\alpha\beta < 0$). 
For the R\"ossler lattice, this can be 
naively achieved by introducing a dispersive-like coupling between
the $c_1$ and $c_2$ variables, replacing (\ref{lattice}) by:
\begin{equation}
\label{crosscoupling}
\dot{C}^i_{t+\tau} = {\cal F}_{\tau}(C^i_t) + \frac{\varepsilon}{2} 
{\small
\pmatrix{1 & -\delta & 0 \cr \delta & 1 & 0 \cr 0 & 0 & 1\cr}}
\left( {\cal F}_{\tau}(C^{i-1}_t) -2 {\cal F}_{\tau}(C^{i}_t) +
{\cal F}_{\tau}(C^{i+1}_t) \right),
\end{equation}
where $\delta$ is a new parameter controlling the dispersive part
of the coupling.

We studied the dynamic scaling of the phase interface with $\delta=0.4$,
and all other parameters as in the previous section. There is a clear
increase in the phase fluctuations, though without the appearance of
defects, as might be expected from a ``phase turbulence-like'' behavior.
As before, the mean saturated square width scales with $L$ (Fig.~\ref{f3}a),
yielding $D/24\nu \simeq 2.85 \times 10^{-5}$. The growth regime of the phase
interface of a large system quickly reaches the scaling regime 
(\ref{nlgrowth}) characteristic of the KPZE (Fig.~\ref{f3}b). 
Even though the roughening remains weak in absolute terms, the $2/3$ exponent
indicates that the particular choice of coupling made here achieved its goal.
The insert of Fig.~\ref{f3}b shows that the skewness of the interface,
a universal ratio of amplitudes, takes the value expected for the 
one-dimensional KPZE.

The scaling laws for $w$ shown in Fig.~\ref{f3} do not allow the
independent determination of $\nu$ and $D$. From Fig.~\ref{f3}b, using
(\ref{nlgrowth}), one gets $\lambda D^2/4\nu^2 \simeq 1.2 \times 10^{-7}$.
This, together with the value $D/24\nu \simeq 2.85 \times  10^{-5}$ measured
from Fig.~\ref{f3}a, actually provides the following 
estimate:  $|\lambda | \simeq 1.0$. 
There exist several ways of 
completing the estimation of 
the KPZE parameters. Here, as we merely wanted to check the consistency
of the KPZE picture, we limited ourselves to a rough fit of the early-time
($t<300$)) growth with the linear regime (\ref{lingrowth}) (not shown).
This gives $D/\sqrt{2\pi\nu} \simeq 2\times 10^{-4}$.

Finally, we find $\nu \simeq 0.5$ and $D \simeq 3.5 \times 10^{-4}$,
leading to $g \simeq 2.4 \times 10^{-3}$, $L_c \simeq 6 \times 10^4$,
and $t_c \simeq 8 \times 10^7$.
While the value of $L_c$ is reasonable in view of our numerical
results, that of $t_c$ seems too large. One must keep in mind, of course,
that these values are only rough
estimates, especially for $t_c$,
given their large variation with
parameters $\nu$, $D$, and $\lambda$ (cf. Eq.~(\ref{crossover})).
An additional quantitative agreement with the KPZE is provided by the
value of the skewness of the interface in the growth regime, which is
very close to the ``universal'' value for the KPZE \cite{KRUG}.

\section{Discussion}

Extrapolating the results of the numerical experiments reported in this 
work to the infinite-size, infinite-time limit, one may first conclude
that phase interfaces of chains of coupled R\"ossler
systems roughen, even if quantitative agreement with the KPZE is
debatable. This implies the loss of the phase coherence observed
in finite systems. But this roughening is extremely weak, especially
in the case of pure diffusive coupling\footnote{In all the numerical
experiments reported here, the width of the phase interface remained smaller
than $2\pi$.}. Using properties of the KPZE,
one can only expect an even weaker roughening in two space dimensions. 
In particular, a very slow logarithmic variation of the 
saturated width with $L$ during an extremely extended linear regime
should be observed
(the crossover scales can easily be huge, given their variation
with parameters for $d=2$) \cite{LHT,TURBPHA}. 
It is not surprising, then, that no loss of 
coherence could be detected within the size/time  range investigated
in \cite{ROSS}.

As mentioned, the validity of the KPZE as the relevant large-scale 
stochastic description has not been firmly established. While the 
situation is satisfactory in the case of diffusive-dispersive coupling,
there are discrepancies for the purely diffusive case: notably
the estimates for $L_c$ and $t_c$ are inconsistent with the fact that
the system was observed to remain in the linear growth regime for
$L=2^{18}$ and  $t> 10^5$. There are, in our view, two possible reasons
for this. First, our estimates of the parameters of the effective KPZE
might be inaccurate, leading to estimates
for the crossover scales that are 
orders of magnitude away from their actual values.
Indeed, given expressions (\ref{crossover}), the values of 
$t_c$ and $L_c$ can change dramatically even with moderate changes
of $\lambda$, $D$, and $\nu$. Moreover,
$\lambda$ is given by the variation of the interface velocity
 near zero tilt (Eq.~(\ref{tilt}),
a region difficult to probe numerically.
Thus, the ``true'' value of $\lambda$ could be 
extremely small, and consequently, $L_c$ and $t_c$ much larger than
the estimates found here.
Second, and this is probably related to the first point,
the ``effective noise'' could well be very different from (\ref{noise})
\cite{NEWMAN}. 
We stress again 
that the chaotic regime of the R\"ossler system used here is characterized 
by strong amplitude fluctuations (in the $(c_1,c_2)$ plane)  and 
quasi-nil phase fluctuations. Thus, the very strong coherence of the phase 
interface in the case of diffusive coupling should not be surprising.
On the other hand, the cross-coupling term added to introduce dispersion
(Sec.~\ref{s4})
does provide a way of obtaining large phase fluctuations directly
coupled to the local amplitude chaos of the R\"ossler system. 

Finally, we would like to come back to the status and role of the CGLE
in the problem studied here. Even though lattices of coupled R\"ossler
systems do exhibit many of the qualitative features of the CGLE,
their equivalence with a CGLE submitted to some noise cannot be a strict one.
Specifically, there are no phase soft modes in the R\"ossler case 
(the ``gauge invariance'' of the CGLE is broken).
The approximate correspondence between the $(c_1,c_2)$ coordinates
of the R\"ossler system and the complex field $A$ of the CGLE overlooks
the role of the $c_3$ variable. A rough interface must, at every moment, 
include points where $c_3$ experiences a sharp peak (Fig.\ref{f1}b).
The effect of such localized structures might well be the cause
of peculiar properties of the ``effective noise'' in a Langevin-like
description. Since an initially flat interface probably resists the 
appearance of such structures, one can imagine a particularly strong
rigidity of the interface yielding small widths, and, ultimately,
very small values of $|\lambda|$.

For the diffusive-dispersive coupling case, on the other hand,
the equivalent CGLE is expected to be in a phase turbulent 
regime\footnote{The interface velocity changes little with the 
dispersive coupling, so that $\beta\simeq 1.06$. Using $|\lambda|\simeq 1$
and assuming, following the stable CGLE, that $\lambda=2(\beta-\alpha)$ 
---which is roughly verified for the CGLE even in phase turbulence regimes
\cite{TURBPHA}--- we have $1+\alpha\beta <0$ and effective CGLE parameters
in the phase turbulence region.},
which was shown in \cite{TURBPHA} to be itself well
described by the KPZE at large scales. Any additional perturbations,
such as those introduced by the $c_3$ variable,
are not expected to alter significantly this picture, in agreement
with our findings.
 
Even though finer numerical investigations are needed to resolve 
the difficulties encountered above, our work once more points
at the subtleties involved when one tries to build a Langevin description
of  chaotic extended systems.

\begin{ack}
We thank CESUP-UFRGS (Porto Alegre) and CEA (Saclay) for computing 
time allowances. LGB wishes to thank CEA-Saclay, where part of this work
was performed, for hospitality. HC thanks Cray-Brazil and UFRGS for making
his stay in Porto Alegre possible.
\end{ack}

\end{document}